\documentclass[rmp,twocolumn, byrevtex]{revtex4}
\usepackage{graphicx}
\usepackage{dcolumn}
\newcommand{\inassb}{InAs$_{x}$Sb$_{1-x}\,$}
\newcommand{\inassbfive}{InAs$_{0.05}$Sb$_{0.95}\,$}

\begin{document}
\title{Alloying induced degradation of the absorption edge of InAs$_{x}$Sb$_{1-x}$}
\author{Bhavtosh Bansal\footnote{bhavtosh.bansal@gmail.com}, V.
K. Dixit, V. Venkataraman, and H. L. Bhat} \affiliation{Department
of Physics, Indian Institute of Science, Bangalore 560 012,}
\begin{abstract}
(Submitted to Applied Physics Letters November 15, 2006; Accepted
January 29, 2007) \vspace{.2cm}

 \inassb alloys show a strong bowing in
the energy gap, the energy gap of the alloy can be less than the
gap of the two parent compounds. We demonstrate that a consequence
of this alloying is a systematic degradation in the sharpness of
the absorption edge. The alloy disorder induced band-tail (Urbach
tail) characteristics are quantitatively studied for \inassbfive.
\end{abstract}
\maketitle It has long been noticed that the low energy tail
region of the density of states of most, possibly all solids has
an exponential character[\onlinecite{ibmreview,urbach}]. Thus
while one always observes an Urbach tail region in the optical
spectra of even the high purity single crystals of III-V
semiconducting compounds[\onlinecite{old InSb, GaAs:greef,
johnson}], a significant band tailing has traditionally been
associated only with amorphous[\onlinecite{cody}] or heavily doped
semiconductors[\onlinecite{efros,Iribarren}]. In particular, the
effects of alloying induced compositional disorder on the
electronic states of the III-V semiconductors were generally
considered benign. This long held view has undergone a major
revision with the discovery of the nitride based semiconductor
alloys, Ga$_{1-y}$In$_{y}$As$_{1-x}$N$_x$ and Ga$_{1-y}$In$_y$N.
It has become evident that the process of alloying such
``mismatched semiconductors"[\onlinecite{special issue dil
nitrides}] can lead to significant deviations from the virtual
crystal picture, strongly changing the electronic and optical
characteristics of the material. Mismatched alloys can typically
be prepared only in dilute concentration due to the large size and
electronegativity difference between the host and substituent
atom. From the perspective of optical measurements, the
compositional disorder leads to photoluminescence linewidths of
many tens of meV, large Stokes' shifts[\onlinecite{IngaN,
bhavtosh_InGaAsN,InPSb}] between the absorption and emission
spectra and an absence of a sharp cut-off in the energy dependent
transmission[\onlinecite{IngaN}].

In this paper we shall discuss the effect of alloying disorder on
the absorption edge on \inassb. These alloys have disorder effects
intermediate between the extremely incompatible GaAs$_{1-x}$N$_x$
and the completely miscible Ga$_{1-y}$Al$_{y}$As. A study of alloy
disorder in these systems is particularly interesting because they
allow for a close to equilibrium growth[\onlinecite{bulk InAsSB
APL}] for a small value of $x$. GaAs$_{1-x}$N$_x$ alloys, on the
other hand, are metastable configurations whose physical
properties show a very strong dependence on sample
history[\onlinecite{klar}]. From the technological point of
view[\onlinecite{antmonides biefeld, Grigorescu, razeghi}], a few
percent of InAs incorporation in InSb leads to a substantial
lowering of the room temperature (RT) energy gap, to within the
mid-infrared 8-12 $\mu m$ transparent atmospheric window. This is
significant because the energy gap of pure InSb---the III-V
semiconductor with the smallest energy gap---is just outside this
region. While previous studies[\onlinecite{Marciniak, InAsSB bulk
optical gap APL, Chen inassb lpe}] have focused only on the energy
gap of the system, alloying also leads to a broadened density of
states and hence a smeared electronic distribution function. This
may offset the advantage of the lowered energy gap.

\begin{figure}[!b]
\begin{center}
\resizebox{!}{7cm} {\includegraphics{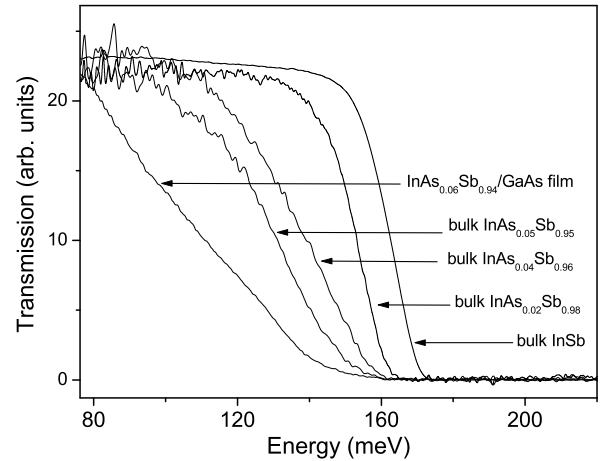}}
\caption{\label{fig:comparative transmission}{RT transmission
spectra for various \inassb samples. Note that the desirable red
shift in the transmission with increasing fraction of InAs is also
systematically accompanied by broadening of the absorption edge.}}
\end{center}
\end{figure}
\noindent {\it Experimental.---}Bulk \inassb samples were grown by
Rotatory Bridgmann method[\onlinecite{bulk InAsSB APL}] and the
$\sim$ 5 $\mu m$ thick heteroepitaxial films on GaAs substrates
were grown by liquid phase epitaxy[\onlinecite{InAsSb/GaAs JAP}].
Since both these crystal growth methods operate close to
equilibrium, the maximum arsenic incorporation was limited to
around 5-6\% if one chose to work with single-phase
single-crystalline samples[\onlinecite{bulk InAsSB APL}]. The
samples were comprehensively characterized for their
structural[\onlinecite{bulk InAsSB APL, InAsSb/GaAs JAP}],
optical[\onlinecite{InAsSB bulk optical gap APL}],
transport[\onlinecite{physica E}], and
magnetotransport[\onlinecite{physica E}] properties and the
results indicated that these were among the best reported so far,
with doping density less than $5\times 10^{16}\textrm{cm}^{-3}$.
Therefore, there is reason to believe that the absorption edge
properties discussed here result from a fundamental limitation of
alloying. The optical transmission spectra were measured using a
Fourier transform infrared spectrometer. The absorption
coefficient $\kappa$ was calculated from the transmission $T$,
using the well-known expression that accounts for multiple
reflections within the sample, $T={(1-R)^2e^{-\kappa d} /[ 1
-R^2e^{-2\kappa d}}]$. Here $d$ is the thickness of the sample and
$R = 0.4$ is the reflection coefficient, assumed to be constant in
the spectral range of measurement. This equation is easily
inverted by substituting $z=e^{-\kappa d}$ and then solving the
quadratic equation in $z$.

The RT transmission from five \inassb samples with progressively
increasing value of $x$ is shown in Fig.1. We observe that the
arsenide alloying induces the expected red-shift in the
gap[\onlinecite{bulk InAsSB APL,InAsSb/GaAs JAP}]. It is also
evident from Fig.1 that a larger arsenic alloying implies a
progressive softening of the absorption edge, as the energy spread
of the band-tail states grows with growing disorder.

To further investigate this band tailing, we shall focus on
\inassbfive$\,$    sample which has the highest alloy composition
among the bulk samples. Fig.2 shows the different parts of the
absorption curve measured on \inassbfive. We can clearly
distinguish three separate regions corresponding to (a) the band
to band absorption[\onlinecite{InAsSB bulk optical gap APL}] where
the absorption coefficient $\kappa \propto [h\nu - E_g(T)]^{1/2}$.
Here $E_g(T)$ is the value of the energy gap at the given
temperature $T$ and $h\nu$ is the incident photon energy, (b) the
Urbach edge absorption region where the absorption coefficient has
an exponential dependence on energy (note the semilog scale in
Fig.2), and (c) the free carrier absorption
region[\onlinecite{InAsSB bulk optical gap APL}] where the
magnitude of the absorption curve increases with wavelength with a
functional form that is dependent on the nature of scattering
potentials[\onlinecite{footnote1}].

\begin{figure}[!t]
\begin{center}
\resizebox{!}{7cm} {\includegraphics{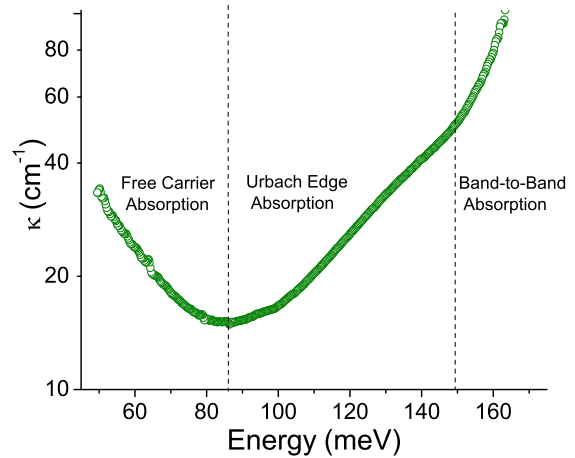}}
\caption{\label{fig:bulk_RT_absorption}{ RT absorption spectrum
for \inassbfive. The spectrum can be separated into three distinct
regions corresponding the to free carrier absorption at smaller
energies, the Urbach edge absorption due to the band-tail states
and finally the direct interband absorption across the energy gap.
}}
\end{center}
\end{figure}
Fig.3(a) highlights the behavior of the exponential Urbach edge
portion of the absorption coefficient for bulk \inassbfive  at
different temperatures between 300K and 433K. Different
exponential curves when extrapolated to higher energies converge
to a single point, the Urbach focus[\onlinecite{cody}]. Hence we
observe that the well-known description of the Urbach edge
developed for amorphous semiconductors seems to also describe the
compositional disorder of mismatched III-V alloys. The exponential
absorption coefficients measured at different temperatures can be
phenomenologically described by the following single
equation[\onlinecite{cody}]:
\begin{equation}\label{eqn:Urbach absorption def}
\kappa(h\nu, T)= \kappa_0 \exp[(h\nu -E_F)/E_U(T)],
\end{equation}
with a  temperature independent $\kappa_0$ and $E_F$. $E_U(T)$ in
the above equation is the temperature dependent (inverse) slope of
the exponential absorption coefficient and is a measure of the
width of the Urbach tail. Henceforth, $E_U(T)$ will be referred to
as the Urbach parameter.

Optical absorption occurs on a timescale much smaller than the
typical phonon frequencies. One may therefore, in the
Born-Oppenheimer spirit, argue that during a direct interband
transition, an electron gets only a snapshot view of its
surroundings and on this short time scale, the dynamically
vibrating crystal is just a distorted arrangement of static
ions[\onlinecite{ibmreview}]. Thus in the context of optical
absorption processes, thermal and structural disorders can be
considered equivalent if the polaronic effects are small. The
Urbach parameter is then expressed[\onlinecite{cody}] as
$E_U=X+Y(T)$, a simple addition of the thermal [$Y(T)$] and
temperature independent structural disorder [$X$] components.

Fig.3(b) shows that the Urbach parameter is a monotonically
increasing function of temperature. Furthermore, Fig.3(c), where
we have plotted the relationship between the energy gap and the
Urbach parameter, suggests a strong correlation between the
temperature dependent changes in the two seemingly unrelated
quantities. Therefore, the temperature dependent contribution is
approximately estimated by assuming the same functional form for
$Y(T)$ as that used to parameterize the change in the energy gap
with temperature. One may[\onlinecite{cody}], for example assume
$Y(T)={Y_0/ [\exp(\theta/T)-1}]$ or even, in our opinion, the
Varshni type relationship, $Y(T)={\tilde{Y_0}T^2/[\tilde{\theta} +
T]}$. Using the Bose-Einstein type relationship, we get [Fig.3(b)]
the value of $\theta\approx 1850K$ and the disorder parameter
$X\approx 35$meV. Note that the value of $\theta$ is not in
agreement with what was earlier inferred for a much larger
temperature range[\onlinecite{InAsSB bulk optical gap APL}]. This
is because the linear relationship between the energy gap and the
temperature dependence of the Urbach parameter is only
approximate. In particular, the temperature dependent
renormalization of the energy gap has a significant thermal
expansion contribution that would not play any obvious role in
changing the width of the Urbach tail. Thus although it is
difficult to make physical sense of the quantity $\theta$, the fit
does give an estimate of the disorder parameter $X$. The
limitations a single temperature $\theta$ to describe the in terms
of say a Debye or Einstein temperature) is has been extensively
discussed in the papers by P\"assler[\onlinecite{passler}] in
context of the temperature dependence of the energy gap.

The universality of the Urbach tail phenomenon across material
systems makes the study of the disorder parameter $X$ particularly
interesting, providing an absolute scale for quantifying the
disorder in a particular sample. It can play the same role for
highly mismatched semiconductor alloys as the scattering time
(mobility) does in less disordered materials.

\begin{figure}[!t]
\begin{center}
\resizebox{!}{10cm} {\includegraphics{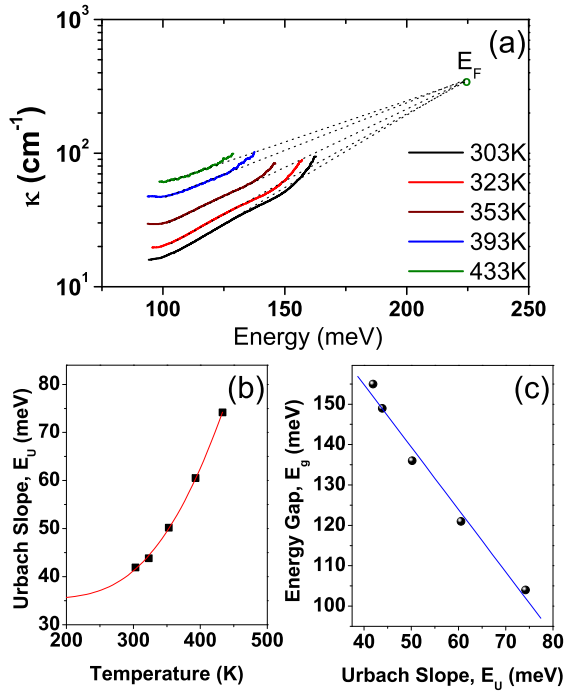}}
\caption{\label{fig:Urbach focus}{ (a) Urbach edge contribution to
the absorption coefficient $\kappa$ at different temperatures for
bulk \inassbfive. All the curves extrapolate to meet at a single
point, denoted by the Urbach focus, $E_F$ at (225 meV, 250
cm$^{-1}$). (b)(squares) Temperature dependence of the slope of
the Urbach edge measured between 303K and 433K. (solid line) Fit
to the relationship described in the text, $\theta=1850K$ and $X=
35$ meV (c) (circles) Relationship between the energy gap,
E$_g$(T) and the Urbach parameter, E$_U$(T). The best straight
line fit has a slope, $G=\Delta E_g/\Delta E_U =-1.55$ }}
\end{center}
\end{figure}
The inferred values of $X\sim 30$ meV and the RT Urbach parameter
$E_U(300K)=42$ meV are intermediate between that observed in high
purity crystalline[\onlinecite{GaAs:greef}] solids ($X< 10$meV)
and that it amorphous[\onlinecite{cody}] materials ($X>50$meV).
Among the few other similar studies which exist for III-V
semiconductor alloys, Reihlen, et al.[\onlinecite{InPSb}] had
measured the RT Urbach parameter of InP$_{1-x}$Sb$_{x}$ to be
between 10meV at small and 50meV at larger alloy compositions.
These results highlight the expected similarity of the two alloy
systems. Studies on In$_y$Ga$_{1-y}$N alloys, on the other hand,
have shown a much larger role of compositional disorder. Bayliss,
et al.[\onlinecite{IngaN}] found the RT Urbach parameter
$E_U(300K)$ in In$_y$Ga$_{1-y}$N to be between 50 and 200 meV and
Han, et al.[\onlinecite{han}] measured the same to be as large as
850 meV in In$_{0.13}$Ga$_{0.87}$N. Such large values may be
attributed to the immiscibility of the two parent
compounds[\onlinecite{han}]. Interestingly, the studies in Refs.
\onlinecite{IngaN} and \onlinecite{InPSb} also correlate the
Urbach parameter with the Stokes' shift. For comparison, very
heavily ($10^{19}$cm$^{-3}$) doped GaAs has an Urbach parameter of
about 20meV at RT[\onlinecite{Iribarren}]. A value of about 90 meV
was obtained for the structural disorder parameter in
polycrystalline InN[\onlinecite{InN}].

Finally, we shall discuss the physical interpretation of the
Urbach focus. Following the theoretical analysis of Abe and
Toyozawa, Cody[\onlinecite{cody}] has suggested that the Urbach
parameter, the Urbach focus and the energy gap, may be related by
$E_g(T)=E_F+GE_U(T)$. Here $G=\Delta E_g/\Delta E_U$. A least
squares fit to our data [Fig.3(c)] yields a slope $G=-1.55$,
although the quality of fit is not very good. The value of the
Urbach focus inferred from the above relation is found to be
between $0.22$ and $0.21$ eV for measurements at different
temperatures and is in reasonably good agreement with the
convergence of the extrapolated exponential absorption spectra at
$0.225$ eV. The Urbach focus in amorphous semiconductors is
identified as the effective direct energy gap of the
(disorder-free hypothetical) virtual crystal[\onlinecite{cody}].
This is both difficult to physically visualize, because of the
topological disorder in real amorphous materials, and its value of
$2.2$ eV in amorphous silicon is not readily related to any other
physically measured quantity in the system. The virtual crystal of
a compositionally disordered solid, on the other hand, is much
easier to visualize. Based on the above equation, one would expect
the difference of the Urbach focus energy and the zero temperature
gap of the real crystal to be equal to the disorder parameter
responsible for the energy gap bowing[\onlinecite{footnote2}]. But
our inferred value of the Urbach focus, at about 0.22 eV, is
closer to the zero temperature gap in this
sample[\onlinecite{InAsSB bulk optical gap APL}] itself,
indicating that the above argument is at best qualitative.

\noindent {\it Conclusions.---}We have studied the optical
absorption edge in crystalline \inassb alloys grown by equilibrium
technique. The absorption edge was found to be sensitive to the
alloy composition, alloying led to a systematic broadening of the
absorption edge. We further studied the Urbach edge's contribution
to the absorption coefficient in bulk \inassbfive and found that
the temperature dependent absorption coefficient curves could be
extrapolated to a single point, the Urbach focus occurring at
about 225 meV. The temperature dependence of the Urbach slope was
modelled to infer a structural disorder energy of about 30meV for
\inassbfive. As expected, this value is intermediate between what
is observed in high quality compound semiconductors and heavily
disordered amorphous solids. The universal nature of the Urbach
edge suggests that the band tail disorder parameter can be used as
a useful measure of sample quality, not just in amorphous solids
but also in disordered III-V semiconductor alloys.

One of the authors (VKD) is presently at the Semiconductor Laser
Section, Solid State Laser Division, Raja Ramanna Centre for
Advanced Technology, Indore 452013, M. P., India.

\end{document}